\documentclass[12pt,preprint]{aastex}

\usepackage{graphicx}
\usepackage{amsmath}

\begin{document}

\title{Three-Dimensional Relativistic Magnetohydrodynamic Simulations of Current-Driven Instability. III. Rotating Relativistic Jets}

\author{
Yosuke Mizuno\altaffilmark{1,2,3}, Yuri Lyubarsky\altaffilmark{4}, Ken-Ichi
Nishikawa\altaffilmark{2,3}, and Philip E. Hardee\altaffilmark{5}}

\altaffiltext{1}{Institute of Astronomy, National Tsing-Hua University
, No. 101, Sec. 2, Kuang-Fu Road., Hsinchu, Taiwan 30013, R.O.C.; mizuno@phys.nthu.edu.tw}
\altaffiltext{2}{Center for Space Plasma and Aeronomic Research, University of Alabama in Huntsville, NSSTC, Huntsville, AL 35805, USA}
\altaffiltext{3}{National Space Science and Technology Center,
VP62, Huntsville, AL 35805, USA}
\altaffiltext{4}{Physics Department, Ben-Gurion University, Beer-Sheva 84105, Israel}
\altaffiltext{5}{Department of
Physics and Astronomy, The University of Alabama, Tuscaloosa, AL
35487, USA}

\shorttitle{3D RMHD Simulations of CD Instability III}
\shortauthors{Mizuno et al.}

\begin{abstract}
We have investigated the influence of jet rotation and differential motion on the linear and nonlinear development of the current-driven (CD) kink instability of  force-free helical magnetic equilibria via three-dimensional relativistic magnetohydrodynamic simulations. In this study, we follow the temporal development within a periodic computational box. Displacement of the initial helical magnetic field leads to the growth of the CD kink instability. We find that, in accord with linear stability theory, the development of the instability depends on the lateral distribution of the poloidal magnetic field. If the poloidal field significantly decreases outwards from the axis, the initial small perturbations grow strongly, and if multiple wavelengths are excited non-linear interaction eventually disrupts the initial cylindrical configuration. When the profile of the poloidal field is shallow, the instability develops slowly and eventually saturates. We briefly discuss implications of our findings for Poynting dominated jets.
\end{abstract}

\section{Introduction}

Astronomical jets are seen in a wide variety of astrophysical objects ranging from small-scale protostellar jets (e.g., Reipurth \& Bally 2001) to large-scale extragalactic jets. The jets from black hole binary systems (microquasars; e.g., Mirabel \& Rodoriguez 1999), active galactic nuclei (AGNs; e.g., Urry \& Padovani 1995; Ferrari 1998; Meier et al. 2001), and gamma-ray bursts (GRBs; e.g., Zhang \& M\'{e}sz\'{a}ros 2004; Piran 2005; M\'{e}sz\'{a}ros 2006) have relativistic speed. Synchrotron emission and rotation measures indicate that magnetic fields are a ubiquitous element in jets (e.g., Asada et al. 2002; Gabuzda et al. 2004).

The deepest parts of AGN jets are explored by studying the properties of blazars. Tremendous progress has been made during the past decade on reproducing the spectral energy distributions (SEDs) of blazars. Fermi/LAT has filled in the frequency gap between coverage provided by X-ray satellites such as Chandra and Suzaku, and that provided by ground based Cherenkov arrays  such as Magic, HESS, and VERITAS. Multi-wavelength observations of blazars probe the structure of the blazar zone (e.g., Marscher et al. 2008, 2010; Abdo et al. 2010). VLBI proper motions supply an estimate of Lorentz factors that link observed timescales to intrinsic timescales, and the distance between the central engine and the mm radio core. The observed rotation of the optical polarization vector in several sources suggests helical twisting of both the flow and the magnetic field between the central engine and the radio core. Radio and optical flux, polarization, and timing information along with proper motions, have made it possible to construct likely scenarios for the dynamics and location of emission in the various different wavebands. Nevertheless, much remains to be understood.

It is thought that jets are powered and collimated by magnetohydrodynamic (MHD) processes (Lovelace 1976; Blandford 1976, 2000; Blandford \& Znajek 1977). General relativistic magnetohydrodynamic (GRMHD) simulations of jet generation (e.g., De Villiers et al. 2003, 2005; Hawley \& Krolik 2006, McKinney \& Gammie 2004; McKinney 2006; McKinney \& Blandford 2009; Beckwith et al. 2008; Hardee et al. 2007; Komissarov \& Barkov 2009; Penna et al. 2010) show development of magneto-rotational instability (MRI) (Balbus \& Hawley 1998) and angular momentum transfer in the accretion disk, leading to diffusion of matter and magnetic field inwards, and unsteady outflows near a centrifugally supported funnel wall. In general, GRMHD simulations with spinning black holes indicate jet production consisting of a Poynting dominated (in the sense that the energy is transferred predominantly by electromagnetic field, i.e. the Poynting flux exceeds the plasma energy flux), high Lorentz factor spine, and a matter dominated, mildly relativistic sheath with $v < c$ possibly embedded in a lower speed, $v << c$, disk/coronal wind.  Note however, the that the Lorentz factor can achieve a maximum in a tenuous boundary layer between a Poynting-flux spine and a dense lower speed kinetically dominated sheath (Mizuno et al. 2008; Aloy \& Mimica 2008; Zenitani et al. 2010a).

The key problem in all models of this sort is the conversion of electro-magnetic energy into plasma energy. Gradual acceleration by magnetic forces (Beskin \& Nokhrina 2006; Komissarov et al 2007, 2009, 2010; Tchekhovskoy et al. 2009, 2010; Lyubarsky 2009, 2010a, 2011; Granot et al. 2011; Lyutikov 2011; Granot 2012a,b) as well as dissipation of alternating magnetic fields (Spruit et al. 2001; Drenkhan 2002, Drenkhahn \& Spruit 2002;  Lyubarsky 2010b; McKinney \& Uzdensky 2012) have been suggested as possible energy conversion mechanisms. On the other hand, the large scale magnetic field may dissipate if the regular magnetic structure is destroyed as a result of a global MHD instability, the kink instability being the most plausible candidate (Lyubarskii 1992, 1999; Eichler 1993; Spruit et al 1997; Begelman 1998; Giannios \& Spruit 2006).

The toroidal magnetic field ($B_{\phi}$) produced in outflows from rotating bodies (neutron stars, black holes and accretion disks) becomes dominant in the far zone because the poloidal field ($B_{p}$) falls off faster with expansion and distance. In configurations with strong toroidal magnetic field, the current-driven (CD) kink mode is unstable. This instability excites large-scale helical motions that can strongly distort or even disrupt the system thus triggering violent magnetic dissipation. This instability is common in many configurations containing strong toroidal magnetic fields, e.g., coronal mass ejection (erupting magnetic flux ropes) from the sun (e.g., Fan 2005; Birn et al. 2006; Inoue \& Kusano 2006), stellar interiors (Bonanno \& Urpin 2011,2012) and pulsar wind nebulae (Begelman 1998; Mizuno et al. 2011a). For static cylindrical equilibria, the well-known Kruskal-Shafranov criterion indicates that the instability develops if the length of the plasma column, $\ell$, is long enough for the field lines to go around the cylinder at least once, i.e., $|B_{p}/B_{\phi}| < \ell/ 2 \pi R$ (e.g., Bateman 1978). In relativistic jets, rotation and velocity shear significantly affect the instability criterion (Istomin \& Pariev 1994, 1996;  Lyubarskii 1999; Tomimatsu et al. 2001; Narayan et al. 2009).

It is known from the field of thermonuclear fusion that the nonlinear development of the kink instability is different for external and internal kink modes (e.g. Bateman 1978). The external instability develops in a current carrying plasma filament surrounded by a vacuum magnetic field; this instability is disruptive. In space conditions, such a configuration is not relevant because some amount of plasma is present everywhere and the magnetic field is frozen into this plasma. In this case the so-called internal kink instability develops. The internal kink mode develops inside a resonance surface on which helicity of the magnetic field lines matches that of the eigenmode. In dissipation-less MHD this instability is known to saturate if the toroidal field is small compared to the poloidal one (e.g. Bateman 1978). Finite resistivity makes the internal kink mode disruptive. However, the corresponding time-scale, although short in laboratory devices, significantly exceeds the Alfv\'en time-scale and is too long to disrupt a fast outflow. On the other hand, the toroidal field is large in astrophysical magnetized outflows and beyond the Alfv\'en surface the toroidal field exceeds the poloidal field. Therefore the question of whether the kink instability is able to disrupt the jet's structure and trigger violent magnetic dissipation remains open.

Only fully 3D simulations allow proper study of the kink instability in the non-linear stage. At this time only a few 3D simulations of relativistic, magnetized jets have been performed, and the results are still controversial. McKinney \& Blandford (2009) simulated the generation and propagation of a relativistic highly magnetized jet and found no significant development of non-axisymmetric instabilities. On the contrary, Mignone et al.\ (2010) found strong wiggling of the jet. The discrepancy may be attributed to the difference in the setup.  Mignone et al.\ (2010) assumed that the magnetic field was purely toroidal, whereas McKinney \& Blandford (2009) simulated  jet launching from a rotating magnetized configuration and their jet contained both toroidal and poloidal magnetic field components.

In a previous paper (Mizuno et al. 2009), we considered helically magnetized static plasma columns (or more generally rigidly moving flows considered in the proper reference frame). Simulation results showed that the initial configuration is strongly distorted but not disrupted by the CD kink instability. We also investigated the influence of a velocity shear surface on the linear and non-linear development of CD kink instability of force-free helical magnetic equilibria in 3D and found that helically distorted density structure propagates along the jet with speed and flow structure dependent on the radius of the velocity shear surface relative to the characteristic radius of the helically twisted force-free magnetic field (Mizuno et al. 2011b). Recently O'Neill et al. (2012) have also investigated the CD kink instability in a local, co-moving frame using the 3D relativistic MHD module within the Athena code (Beckwith \& Stone 2011). Their results for a static, force-free configuration are consistent with ours. They also studied rotating jets with a purely toroidal magnetic field, the hoop stress being balanced by the pressure and centrifugal force, and found that in this case, the instability brings the system into a turbulent state.

In this paper, we investigate the nonlinear behavior of the CD kink instability in a differentially rotating jet. As in our previous studies, we perform simulations in a reference frame co-moving with the jet so that the plasma velocities in our frame are only mildly relativistic (in the general case of differentially moving flow, there is no  frame of reference where the velocities vanish). As our goal is to investigate step by step the role of different factors in the development of the instability, here we concentrate on the differential rotation and neglect for a while the sideways expansion of the jet even though the last could presumably slow down or even suppress the instability. These effects deserve a separate study. For a while we assume that the jet expands slowly enough. We consider configurations close to the force-free equilibrium of a rotating jet, in which poloidal and toroidal magnetic fields are comparable and the transverse equilibrium is maintained by a balance between the hoop stress, pressure of the poloidal field and the electric force where the centrifugal force and the gas pressure remain small. 

Our setup is motivated by an analysis of the structure of Poynting dominated jets (Lyubarsky 2009). Even though the magnetic field in the jet is predominantly toroidal, the poloidal field  cannot in general be neglected because one has in fact to compare the fields in the comoving frame where the toroidal field is much less than in the lab frame.  In cylindrically equilibrium configurations, the poloidal and toroidal fields are comparable in the comoving frame. The jet structure relaxes to a locally equilibrium configuration if the jet is narrow enough so that the Alfv\'en crossing time is less than the proper propagation time.  An important point is that only such jets could be subject to the kink instability because the characteristic instability time-scale is larger than the Alfv\'en crossing time. Therefore stability analysis is relevant only for jets with non-negligible poloidal field. The poloidal field can be neglected only if the jet expands rapidly enough so that the Alfv\'en crossing time exceeds the proper jet propagation time; then the jet is unable to relax to an equilibrium configuration and therefore the hoop stress remains unbalanced. Such jets can indeed be considered as composed from concentric magnetic loops; however, no global MHD instabilities can develop in causally disconnected  flows therefore we do not consider such configurations here. 

Note that causality considerations (in terms of Alfv\'en crossing time) are of primary importance even in non-relativistic jets, see analysis by Moll et al. (2008). An important difference is that beyond the Alfv\'en point, a non-relativistic jet could not be in transverse equilibrium because the toroidal field exceeds the poloidal one and the hoop stress is not counterbalanced by the electric force as in relativistic flows. In non-relativistic jets, the flow reaches approximate energy equipartition already near the Alfv\'en point and beyond this point, where the kink instability could develop, most of the energy is already transferred to the plasma (Moll et al. 2008, Moll 2009). It is a specific property of relativistic flows that well beyond the Alfv\'en point, the flow could remain Poynting dominated and the transverse structure of the flow could be close to that in cylindrically equilibrium magnetic configurations.

In this paper, we focus on the temporal development of CD kink instability in a differentially rotating relativistic jet with periodic box. In this situation, we follow the evolution of a few wavelengths of instability in a reference frame within a small simulation domain. Many previous relativistic and non-relativistic MHD simulations of jet formation and propagation (e.g., Nakamura \& Meier 2004; Nakamura et al. 2007; Moll et al. 2008; Moll 2009; McKinney \& Blandford 2009; Mignone et al. 2010) have followed the spatial properties of development of kink instability in the jet. Even though only such simulations could take into account properly real conditions, it is difficult to figure out a role of different factors and therefore to gain physical insights. Therefore we choose the investigation of temporal development of instabilities in an infinite cylindrical jet. At this step, we concentrate on the role of the differential rotation, which introduces an important physical factor: the net electric force, which opposes to the hoop stress. Therefore differentially rotating jets could be in the transverse equilibrium even if the transverse gradient of the poloidal field is zero. The linear stability analysis shows (Istomin \& Pariev 1994, 1996; Lyubarskii 1999) that the instability growth rate goes to zero when the transverse gradient of the poloidal field vanishes (but the poloidal field has to be comparable with the toroidal field in the comoving frame). Here we investigate dependence of the non-linear development of the instability on the transverse structure of the jet. We demonstrate that the instability could become disruptive only if the poloidal field significantly decreases outwards from the axis. When the profile of the poloidal field is shallow, the instability develops slowly and eventually saturates.  

We describe the numerical method and setup used for our simulations in \S 2, present our results in \S 3, in \S 4 compare our results to instability expectations and conclude.

\section{Numerical Method and Setup}

In order to study time evolution of the CD kink instability in the relativistic MHD (RMHD) regime, we use the 3D GRMHD code ``RAISHIN'' in Cartesian coordinates. RAISHIN is based on a $3+1$ formalism of the general relativistic conservation laws of particle number and energy-momentum, Maxwell's equations, and Ohm's law with no electrical resistance (ideal MHD condition) in a curved spacetime (Mizuno et al.\ 2006).  In the RAISHIN code, a conservative, high-resolution shock-capturing scheme is employed. The numerical fluxes are calculated using the HLL approximate Riemann solver, and flux-interpolated constrained transport (flux-CT) is used to maintain a divergence-free magnetic field\footnote{Constrained transport schemes are used to maintain divergence-free magnetic fields in the RAISHIN code. This scheme requires the magnetic field to be defined at the cell interfaces. On the other hand, conservative, high-resolution shock capturing schemes (Godonov-type schemes) for conservation laws require the variables to be defined at the cell center. In order to combine variables defined at these different positions, the magnetic fields at the cell interfaces are interpolated to the cell center and as a result the scheme becomes
non-conservative even though we solve the conservation laws (Komissarov 1999).}. We have used the MC slope-limiter scheme (second order) for reconstruction in the simulations performed in this paper.
The RAISHIN code performs special relativistic calculations in Minkowski spacetime by choosing the appropriate metric. The RAISHIN code has proven to be accurate to second order and has passed a number of numerical tests including highly relativistic cases and highly magnetized cases in both special and general relativity (Mizuno et al.\ 2006, 2011b).

The aim of our simulations is to elucidate the impact of the kink instability on Poynting dominated jets.  The basic picture we keep in our minds is the following.  A rotating black hole is embedded in a large-scale magnetic field supported by currents flowing in the accretion disk. A magnetized wind flows outwards both from the black hole and from the disk but most of the energy is transported along the field lines threading the horizon and, maybe, the inner part of the disk (Blandford 1976; Lovelace 1976; Blandford \& Znajek 1977). The kink instability comes into play when the toroidal field in the comoving frame becomes comparable to the poloidal field. This occurs at distances small compared to the outer disk radius (but beyond the ``light cylinder" radius); the flow in this region may be considered as quasi-cylindrical with the magnetic field having a maximum within some core and gradually decreasing in the lateral direction. No external medium reaches this zone. Therefore
we consider cylindrical magnetic configurations with a well-defined core and extended wings. Our goal is to understand how the kink instability affects the structure and dynamics of the core where most of the magnetic energy is concentrated.

In our simulations we choose a force-free helical magnetic field for the initial configuration (Mizuno et al. 2009).
A force-free configuration is a reasonable choice for the Poynting dominated jet. In general, the force-free equilibrium of a cylindrical rotating jet is described by the equation
\begin{equation}
\frac{\Omega B_{z}}{c^2} {d \over dR} \Omega R^{2} B_{z} =B_{z} {d B_{z} \over dR} +{B_{\phi} \over R} {d B_{\phi} \over dR},
\end{equation}
where $c$ is the speed of light.
In particular we choose the following form for the poloidal ($B_{z}$) field component and the angular velocity of a  magnetic field line ($\Omega$)
\begin{equation}
B_{z}= {B_{0} \over [1+ (R/R_{0})^{2}]^{\alpha}}~,
\end{equation}
\begin{equation}
\Omega=
\left\{ \begin{array}{cl}
 \Omega_{0} & \textrm{if $R \le R_{0}$} \\
 {\Omega_{0} (R_{0}/R)^{\beta}} & \textrm{if $R > R_{0}$}
 \end{array}
\right. ,
\end{equation}
where $R$ is the radial position in cylindrical coordinates normalized by a simulation scale unit $L \equiv 1$, $B_{0}$ is the magnetic field amplitude, $\Omega_0$ is the angular velocity amplitude, $R_{0}$ is the radius of the core, $\alpha$ is a  poloidal field profile parameter, and $\beta$ is an angular velocity profile parameter.
In simulations we choose $R_{0}=(1/4)L$, and $\beta=1$. Then the toroidal component ($B_{\phi}$) of the magnetic field is given by
\begin{equation}
B_{\phi}= -{B_{0} \over R[1+ (R/R_{0})^{2}]^{\alpha}} \sqrt{ \frac{\Omega^{2} R^{4}}{c^2} +
{ R_{0}^{2} [1 + (R/R_{0})^{2}]^{2 \alpha} -R_0^{2} - 2 \alpha R^{2} \over 2 \alpha -1}}~.
\end{equation}
When $\Omega_{0}=0$, the force-free helical magnetic field is the same as that used for a static plasma column (Mizuno et al. 2009).  The pitch, $P \equiv R B_{z} / B_{\phi}$, indicates the magnetic field pitch and smaller $P$ indicates more tightly wrapped magnetic field lines. With our choice for the force-free field, the pitch can be written as
\begin{equation}
P = R^{2} \sqrt{ { 2 \alpha -1 \over (2 \alpha -1) (\Omega R^{2}/c)^2 +
R_{0}^{2}[1 + (R/R_{0})^{2} ]^{2 \alpha} -R_0^{2} -2 \alpha R^{2}} }~.
\end{equation}
If the poloidal field profile parameter $\alpha < 1$,  the pitch $P$ increases and the magnetic field helicity decreases with radius, i.e., the magnetic field is not wrapped as tightly at larger radius. When $\alpha =1$ and $\Omega_{0}=0$, the pitch and magnetic helicity are constant and if $\alpha > 1$, the pitch decreases and the magnetic helicity increases with radius, i.e., the magnetic field is wrapped more tightly at larger radius. At larger $\Omega_{0}$, the pitch is smaller and the magnetic field is wrapped more tightly.

In force-free MHD, the true plasma velocity is undefined. One can define only the drift velocity, ${\bf v}=c{\bf E\times B}/B^2$, whereas the component along the magnetic field remains arbitrary. In the chosen configuration, the poloidal and toroidal components of the drift velocity are given by
\begin{equation}
v_{z}=-{B_{\phi} B_{z} \over B^{2}} \Omega R,
\end{equation}
\begin{equation}
v_{\phi}=(1- {B^{2}_{\phi} \over B^{2}}) \Omega R.
\end{equation}
We note that $\Omega > 0$, $v_{z} >0$ and $B_{z} > 0$ imply $B_{\phi} < 0$ and the toroidal magnetic field component is wrapped in the opposite direction from $v_{\phi}$. The velocity approaches light speed when $\Omega \gg 0$.  We choose the parameters so that $v_z$ is not too close to $c$, which in fact means that we are working in a frame comoving with the jet. Of course, in a differentially moving flow there is no single comoving frame; we chose one where the flow velocities are not highly relativistic. In this frame, the unstable perturbation develops in time. In the lab frame, the perturbation develops in space, and with relativistic time dilation the proper time-scale $t$ is transformed to a spatial scale as $z=c\gamma t$, where $\gamma$ is the Lorentz factor of the jet.

We consider a low gas pressure medium with pressure decreasing radially, similar to the angular velocity decrease with radius, and with $p_{0} =0.01$ (see Eq. 3) in units of $\rho_{0}c^{2}$.  This pressure profile keeps the sound speed below the Alfv\'en speed in the whole simulation region. A density decreasing with the magnetic field strength as $\rho = \rho_{1} B^2$ with $\rho_{1} =6.25 \rho_{0}$ is chosen in order to keep the  Alfv\'{e}n speed high everywhere in the simulation region. The equation of state is that of an ideal gas with $p=(\Gamma -1) \rho e$, where $e$ is the specific internal energy density and the adiabatic index $\Gamma=5/3$.
The specific enthalpy is $h \equiv 1+e/c^{2} +p/\rho c^{2}$. The magnetic field amplitude is $B_{0} =0.8$ in units of  $\sqrt{4\pi\rho_{0}c^{2}}$ leading to a low plasma-$\beta$. The sound speed is $c_{s}/c \equiv (\Gamma p/\rho h)^{1/2}$ and the Alfv\'{e}n speed is given by $v_{A}/c  \equiv [b^{2}/(\rho h +b^{2})]^{1/2}$, where $b$ is the magnetic field measured in the comoving frame, $b^{2}=\mathbf{B}^{2}/\gamma^{2}+(\mathbf{v} \cdot \mathbf{B})^{2}$ (Komissarov 1997; Del Zanna et al. 2007).

In order to investigate the effect of jet rotation, we perform simulations with four different angular velocities, $\Omega_{0}=1$, $2$, $4$, and $6$. Results are compared to those for a static plasma column (no flow, $\Omega_{0}=0$) as a reference.  Radial profiles of the angular velocity, the magnetic field components ($B_{\phi}$, $B_{z}$), the magnetic pitch, P, the velocity components ($v_{z}$, $v_{\phi}$), the sound and Alfv\'{e}n speeds, and the density for the different angular velocity cases with $\alpha=1$ are shown in Figure 1. %
\begin{figure}[h!]
\epsscale{0.8}
\plotone{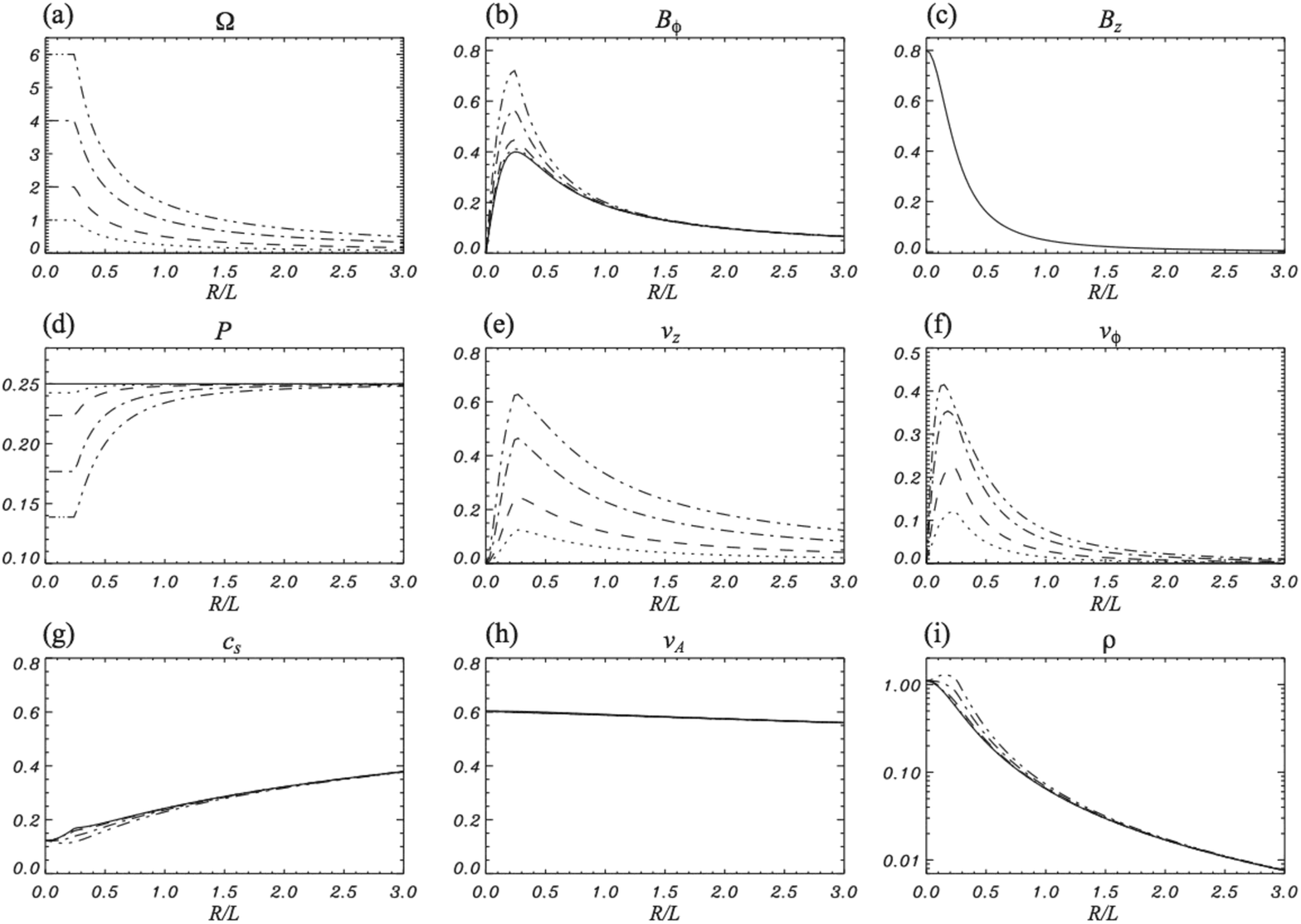}
\caption{Radial profiles of (a) the angular velocity $\Omega$, (b) the toroidal magnetic field $|B_{\phi}|$, (c) the axial magnetic field $B_{z}$, (d) the magnetic pitch $P$, (e) the axial velocity $v_{z}$, (f) the toroidal velocity $v_{\phi}$, (g) the sound speed $c_s/c$, (h) the Alfv\'{e}n speed $v_A/c$, and (i) the density $\rho$ for $\Omega_{0}=0$ (solid line), $1$ (dotted line), $2$ (dashed line), $4$ (dash-dotted line), and $6$ (dash-two-dotted line) cases with $\alpha=1$.
\label{f1}}
\end{figure}
When $\Omega_0$ is increased, the maximum toroidal magnetic field at $R=0.25L$ becomes larger and the magnetic pitch, P, becomes smaller in the core, i.e., the magnetic field is more tightly wrapped. The axial and rotation velocities become faster with larger $\Omega_{0}$ but the rotation velocity remains less than the Alfv\'{e}n speed in all regions. For all cases the sound and Alfv\'en speeds on the axis are $c_{s0} = 0.142~c$ and  $v_{A0} = 0.6~c$.

We also investigate the effect of different magnetic pitch profiles with $\alpha=1$, $0.75$, $0.5$, and $0.35$.  Radial profiles of the angular velocity, the magnetic field components ($B_{\phi}$, $B_{z}$), the magnetic pitch, P, the velocity components ($v_{z}$, $v_{\phi}$), the sound and Alfv\'{e}n speeds, and the density for the different pitch profiles with $\Omega_0=4$ are shown in Figure 2 .
\begin{figure}[h!]
\epsscale{0.8}
\plotone{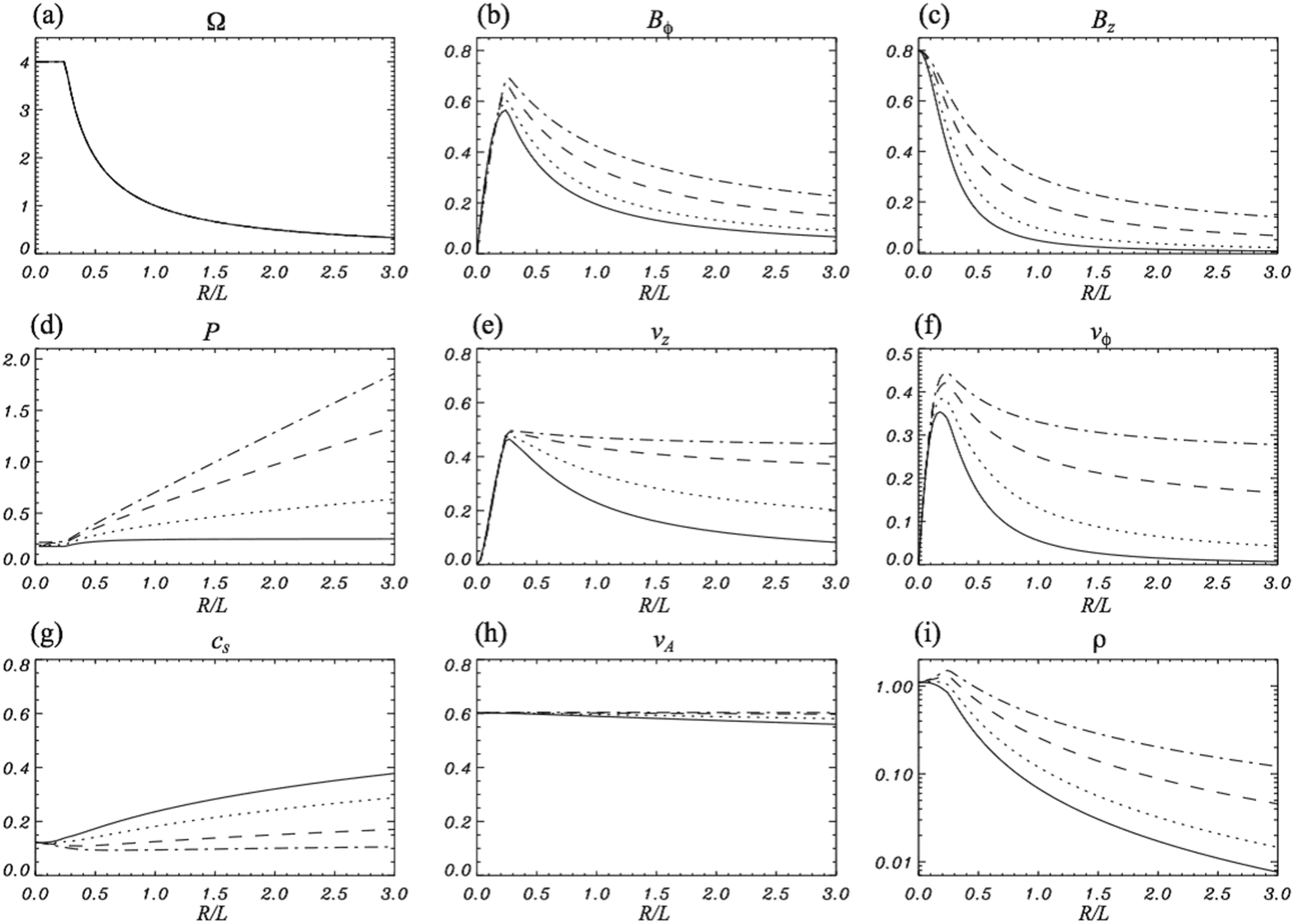}
\caption{Radial profiles of (a) the angular velocity $\Omega$, (b) the toroidal magnetic field $|B_{\phi}|$, (c) the axial magnetic field $B_{z}$, (d) the magnetic pitch $P$, (e) the axial velocity $v_{z}$, (f) the toroidal velocity $v_{\phi}$, (g) the sound speed $c_s/c$, (h) the Alfv\'{e}n speed $v_A/c$, and (i) the density $\rho$ for $\alpha=1$ (solid line), $0.75$ (dotted line), $0.5$ (dashed line), and $0.35$ (dash-dotted line) cases with $\Omega_{0}=4$.
\label{f2}}
\end{figure}
For smaller values of $\alpha$ the magnetic pitch parameter P increases and the magnetic field is less tightly wrapped outside the core. Overall the magnetic field strength increases and declines less rapidly outside the core with decreasing $\alpha$.  The drift velocity also increases and declines less rapidly outside the core with decreasing $\alpha$.  For all cases the sound and Alfv\'en speeds on the axis are $c_{s0} = 0.142~c$ and  $v_{A0} = 0.6~c$.  The Alfv\'{e}n speed is almost independent of $\alpha$. The drift speed is comparable to the Alfv\'en speed outside the core for $\alpha = 0.35$.

The simulation grid is periodic along the axial z direction. The grid is a Cartesian ($x, y, z$) box of size $6L \times 6L \times L_{z}$. $L_{z}$ is the axial grid length. In most cases, we choose the axial grid length, $L_{z}=3L$. The grid resolution is the same in all directions with $\Delta L = L/40$.  In terms of the radius of the jet core, $R_{0}$, the simulation box size is $24R_{0} \times 24R_{0} \times 12R_{0}$ and the allowed axial wavelengths are restricted to $\lambda = 12R_{0}/n \le 12R_{0}$. This simulation grid is the same as that used for case B in Mizuno et al. (2009) with $R_{0} = a$. To investigate multiple wavelength coupling, we performed several simulations with a longer axial simulation box (which are labeled as cases {\it lg}). The grid length for each case is given in Table 1.
\begin{deluxetable}{lccc}
\tablecolumns{7}
\tablewidth{0pc}
\tablecaption{Models and Parameters}
\label{table1}
\tablehead{
\colhead{Case} & \colhead{$\Omega_{0}$} & \colhead{$\alpha$} & \colhead{$L_{z}/L$}
}
\startdata
alp1om0 & 0 & 1.0 & 3 \\
alp1om1 & 1 & 1.0 & 3  \\
alp1om2 & 2 & 1.0 & 3  \\
alp1om4 & 4 & 1.0 & 3  \\
alp1om6 & 6 & 1.0 & 3  \\
alp1om2{\it lg} & 2 & 1.0 & 5  \\
alp1om4{\it lg} & 4 & 1.0 & 4   \\
alp075om4 & 4 & 0.75 & 3 \\
alp050om4 & 4 & 0.5 & 3 \\
alp035om4 & 4 & 0.35 & 3 \\
\enddata
\end{deluxetable}

We impose fixed boundary conditions on the transverse boundaries at $x = y = \pm 3L$ ($\pm 12R_{0}$) to maintain jet rotation. Previously, we investigated the influence of  grid resolution on a static plasma column for four grid resolutions from 20 to 60 computational zones per simulation length unit $L= 8a = 8R_{0}$ (Mizuno et al. 2009). We found that the growth rate depended on grid resolution, but converged to nearly the same value for grid resolutions of $\Delta L=L/40$ and $\Delta L=L/60$.  Thus, in the previous simulations we chose and here use $\Delta L=L/40$ as providing sufficient grid resolution to correctly reproduce the development of the CD kink instability.

To break the symmetry the initial MHD equilibrium configuration is perturbed by a radial velocity profile given by
\begin{equation}
v_{R} /c = {\delta v \over N } \exp \left(-  { R \over R_{p} } \right) \sum^{N}_{n=1} \cos (m
\theta) \sin \left( { \pi n z \over L_{z}}\right)~.
\end{equation}
The amplitude of the perturbation is $\delta v = 0.01$ with radial width $R_{p}= 0.5L$ ($2R_{0}$), and we choose $m=1$ and $N=8$ which excites $n=0.5$, $1$, $1.5$, $2$, $2.5$, $3$, $3.5$, and $4$ kink mode wavelengths.

We use the HLL approximate Riemann solver coupled with a second-order MC slope-limiter reconstruction scheme in the simulations. O'Nell et al. (2012) showed that the choice of Riemann solver scheme can affect the evolution more than the choice of physical parameters. However, they found that only a $\sim 17\%$ kinetic energy difference existed between simulations computed with an HLL solver and an HLLD solver (Mignone et al. 2009)
coupled to a second-order reconstruction scheme for a force-free configuration. This  kinetic energy difference was found to be much less than for non-force free configurations.  Our previous numerical tests when compared to other tests published in the literature indicated that this level of difference can also result from differences in details of otherwise seemingly identical numerical schemes.  Overall we believe that our choice of numerical scheme is adequate for the present force-free configurations.

\section{Results}

\subsection{Dependence on jet rotation}

Figure 3 shows the time evolution of a density isosurface for a static plasma column with $\Omega_{0}=0$ and $\alpha=1$ (alp1om0), where the time, $t$, is in units of $ t_c \equiv R_{0}/c$.
\begin{figure}[h!]
\epsscale{0.65}
\plotone{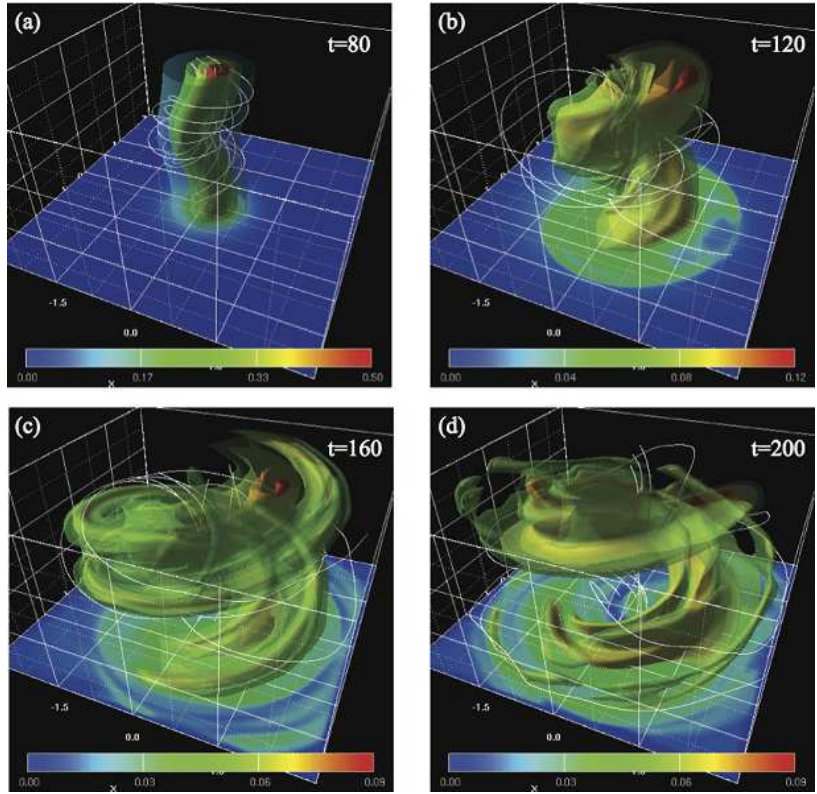}
\caption{Time evolution of three-dimensional density isosurfaces with a transverse slice at $z=0$ for case alp1om0 (static plasma column). The time, $t$, is in units of $t_c = R_0/c$. Color shows the logarithm of the density with solid magnetic field lines. The growth of the CD kink instability leads to a helically twisted magnetic filament wound around the density isosurface associated with the $n=1$ kink mode wavelength. \label{3Drb_alp1om0}}
\end{figure}
The initial conditions are similar to those used in Mizuno et al. (2009) but now with two times stronger initial magnetic field strength and a radially decreasing gas pressure profile.  As seen in Mizuno et al. (2009), displacement of the initial force-free helical magnetic field by growth of the CD kink instability leads to a helically twisted magnetic filament wound around the density isosurface associated with the $n=1$ kink mode wavelength. In the nonlinear phase, helically distorted density structure shows rapid transverse growth and disruption of the high density plasma column. The behavior in the nonlinear phase is quite different from the results shown in Mizuno et al. (2009), where the instability was found to be not disruptive. The difference should be attributed to a different setup. In Mizuno et al (2009), the initial plasma density and pressure were homogeneous so that the system became matter dominated at the periphery where the magnetic field decreases. In the present simulations, we choose the plasma pressure and the density decreasing outward from the axis together with the magnetic field so that the system is magnetically dominated in the whole simulation region.

The time evolution of a density isosurface for $\Omega_{0}=1$ with $\alpha=1$ (alp1om1) is shown in Figure 4.
\begin{figure}[h!]
\epsscale{0.65}
\plotone{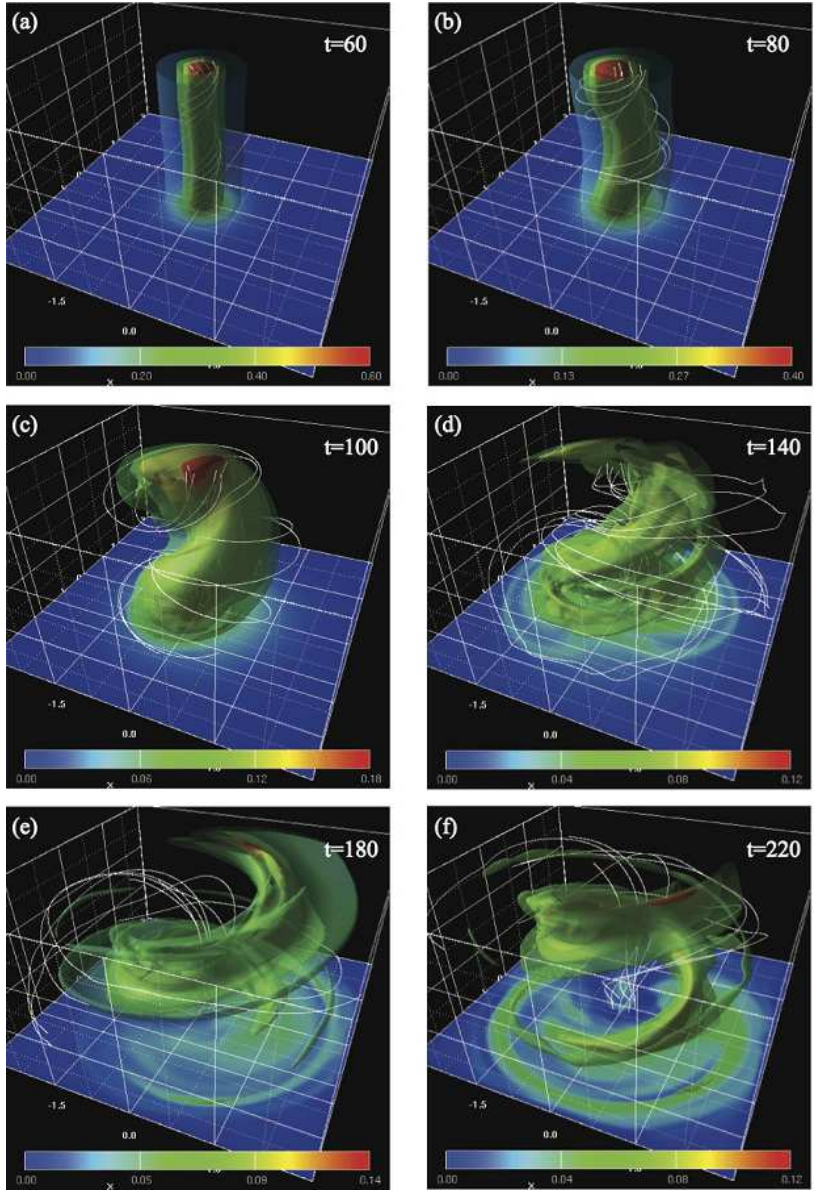}
\caption{Time evolution of three-dimensional density isosurfaces with a transverse slice at $z=0$ for case alp1om1 ($\alpha=1$, $\Omega_0=1$). The time, $t$, is in units of $t_c = R_{0}/c$. Color shows the logarithm of the density with solid magnetic field lines. In the nonlinear phase, helically distorted density structure shows continuous transverse growth and propagates in the flow direction. \label{3Drb_alp1om1}}
\end{figure}
Here the displacement of the initial force-free helical magnetic field resulting from growth of the CD kink instability leads to a helically twisted magnetic filament wound around the density isosurface near the axis associated with the $n=1$ kink mode wavelength. In the nonlinear phase, helically distorted density structure shows continuous transverse growth and propagates in the flow direction (see Fig. 5). The propagation speed of the kink is estimated from the difference in position along z of the high density regions in the xz-plane at $y=0$ at the two times. The estimated speed of the kink is $\sim 0.1c$ and on the order of the initial maximum axial drift velocity of $\sim 0.1c$ (see Figure 1). This propagation continues in the nonlinear phase. The propagation of the kink is reminiscent  of that seen in previous sub-Alfv\'{e}nic CD kink unstable jet flow simulations (Mizuno et al. 2011b).
\begin{figure}[h!]
\epsscale{0.8}
\plotone{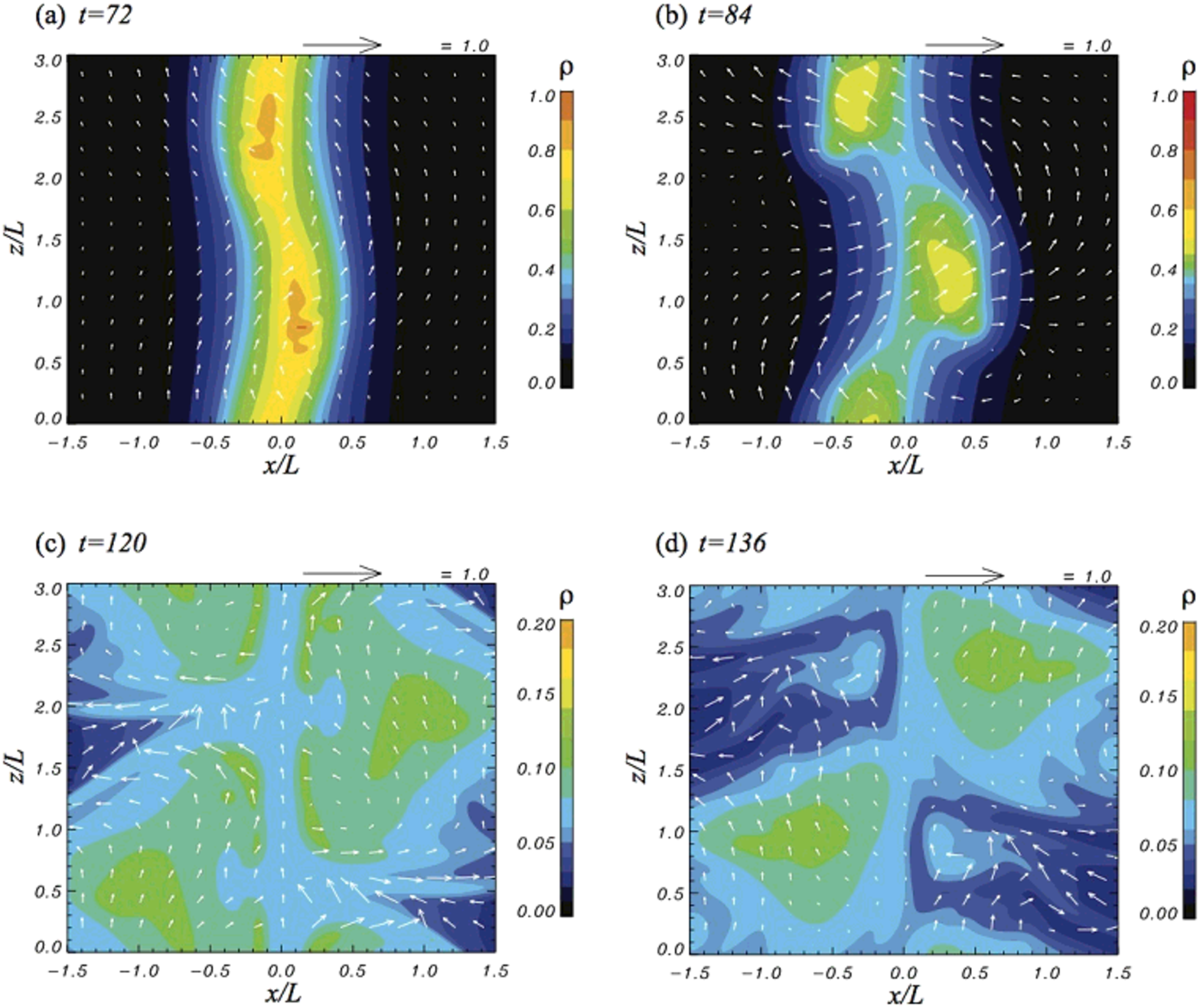}
\caption{Time evolution of two-dimensional images of density in the xz-plane at $y=0$ for case alp1om1 ($\alpha=1$, $\Omega_0=1$). Arrows indicate the poloidal velocity. In this figure, the dependence of instability on the angular rotation is shown. The difference in position of the high density regions indicates the propagation of helically distorted density structure along z axis. \label{2Dxz_ro_alp1om1}}
\end{figure}

In order to investigate the dependence of kink growth on rotation, we have considered four different angular velocities from $\Omega_{0}=0$ to $6$ (see Table 1). In Figure 6 we show the time evolution of the volume-averaged relativistic energies within a cylinder of radius $R/L \le 2.0 ~(R \le 8R_{0})$ as an indicator of the growth of the CD kink instability.
$E_{kin,xy}$ is a volume-averaged kinetic energy transverse to the $z$-axis (Mizuno et al. 2009, 2011),
\begin{equation}
E_{kin,xy}={1 \over V_{b}} \int_{V_b} {\rho v^{2}_{x} + \rho v^{2}_{y} \over 2} dx dy dz,
\end{equation}
and the difference between this value and the initial value ($E_{kin,xy}$ at $t=0$) indicates the radial motion induced by the growth of CD kink instability. $E_{EM}$ is the volume-averaged total relativistic electromagnetic energy written as
\begin{equation}
E_{EM}={1 \over V_{b}} \int_{V_b} {\mathbf{B}^2 + [\mathbf{v}^2 \mathbf{B}^2 - (\mathbf{v} \cdot \mathbf{B})^2] \over 2} dx dy dz.
\end{equation}
\begin{figure}[h!]
\epsscale{0.8}
\plotone{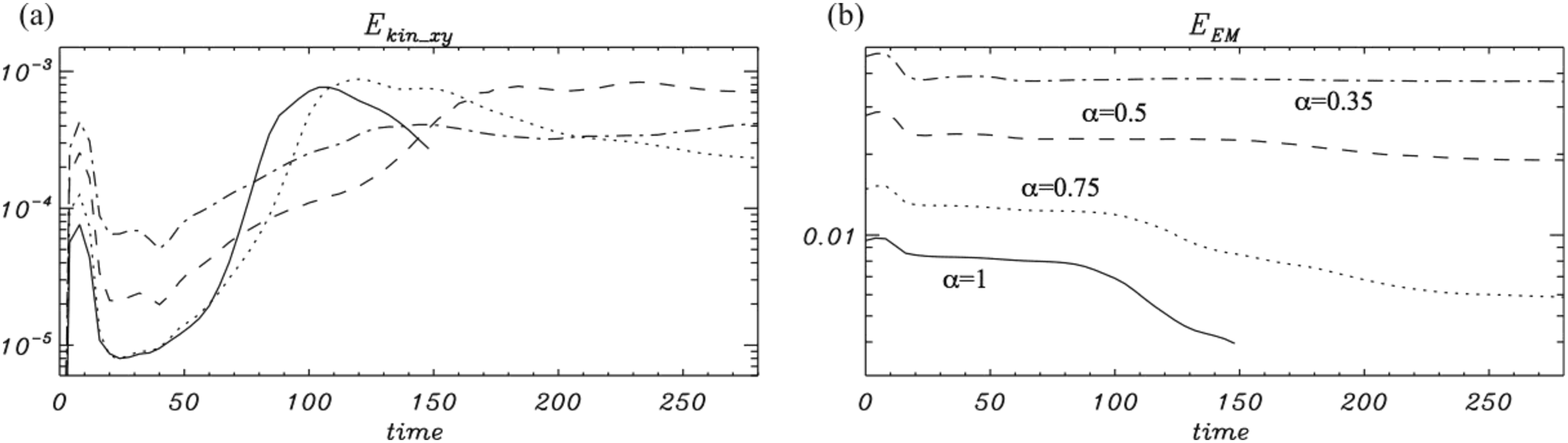}
\caption{Time evolution of volume-averaged (a) $E_{kin,xy}$ and (b) $E_{EM}$ for $\alpha=1.0$ with $\Omega_{0}=0$ (solid line), $1$ (dotted line), $2$ (dashed line), $4$ (dash-dotted line) and $6$ (dash-two-dotted line) within a cylinder of radius $R/L \le 2.0$. $E_{kin,xy}$ is the integrated kinetic energy transverse to the $z$-axis. The initial growth phase is characterized by an exponential increase in $E_{kin,xy}$ to a maximum amplitude followed by a slow decline in the nonlinear phase.
\label{tev_ep1omg1b}}
\end{figure}
The first bump in the kinetic energy curve (at $t<20$), should be attributed to initial relaxation of the system to the cylindrical equilibrium. Recall that the centrifugal and pressure forces at our setup are small but not negligible so that our initial force-free configuration (eqs. 1-4) is not a true equilibrium even though close to it. Therefore a little relaxation occurs for a few Alfv\'en times. Inspection of Fig. 6 shows that the kink instability comes into play at $t>40$, well after the relaxation finishes.

The evolution of $E_{kin,xy}$ in the simulations shows that a characteristic time for the development of the instability is about a dozen Alfv\'en crossing times, i.e., exponential growth from a minimum near $t \sim 40$ to a maximum at $t\lesssim 120$. This agrees with the general estimate of the maximum instability growth rate,
\begin{equation}
\Gamma_{\rm max}\approx 0.1 v_{A}/R_0,
\end{equation}
where $R_0$ is the core radius. The exact coefficient depends on the transverse distribution of the parameters; specifically for constant pitch with $\alpha=1$ and uniform density, Appl et al. (2000) found $\Gamma_{\rm max} = 0.133 v_{A0}/R_0$. Note that this estimate is made in the frame moving with the kink; for a relativistic jet, the growth rate of the instability in the lab frame decreases by the Lorentz factor (Lyubarskii 1999).
 The evolution of $E_{EM}$ is opposite to the time evolution of the kinetic energy as it should be because the CD kink instability develops at the expense of the magnetic energy.

In the internal kink instability, unstable modes match the helicity of the magnetic field lines so that the wavelength of unstable perturbations is determined by the magnetic pinch, $\lambda=2\pi P$ (e.g. Bateman 1978). This condition remains valid also in relativistic flows (Lyubarsky 1999)\footnote{In static magnetically dominated configurations, the hoop stress is balanced by the pressure of the poloidal field therefore the poloidal and toroidal components of the magnetic field should be comparable, at least in the core where the field is maximal (at the periphery, the toroidal field itself could become force-free, $B_{\phi}\propto 1/r$, therefore the pitch could be small there, like in the configuration considered by Appl et al (2000)).  Therefore the pitch, as well as the wavelength of the unstable perturbations, could not be less than the core radius in this case. In relativistic flows, the toroidal field could be larger than the poloidal because the hoop stress could be nearly balanced by the electric force; then the pitch could be less than the jet radius. In highly relativistic jets, the toroidal field significantly exceeds the poloidal one in the lab frame; therefore the wavelength of unstable perturbations are small, in the lab frame, as compared with the jet radius. This may be thought of as a relativistic contraction of unstable loops.}. Our simulations indeed show a decrease of the unstable wavelength with decreasing pitch.

Figure 7 shows three dimensional density isosurfaces and Figure 8 shows two dimensional images of the xz-plane at $y=0$ for the constant pitch cases for the angular velocity amplitudes $\Omega_0=2$, $4$, and $6$.
\begin{figure}[hp!]
\epsscale{0.75}
\plotone{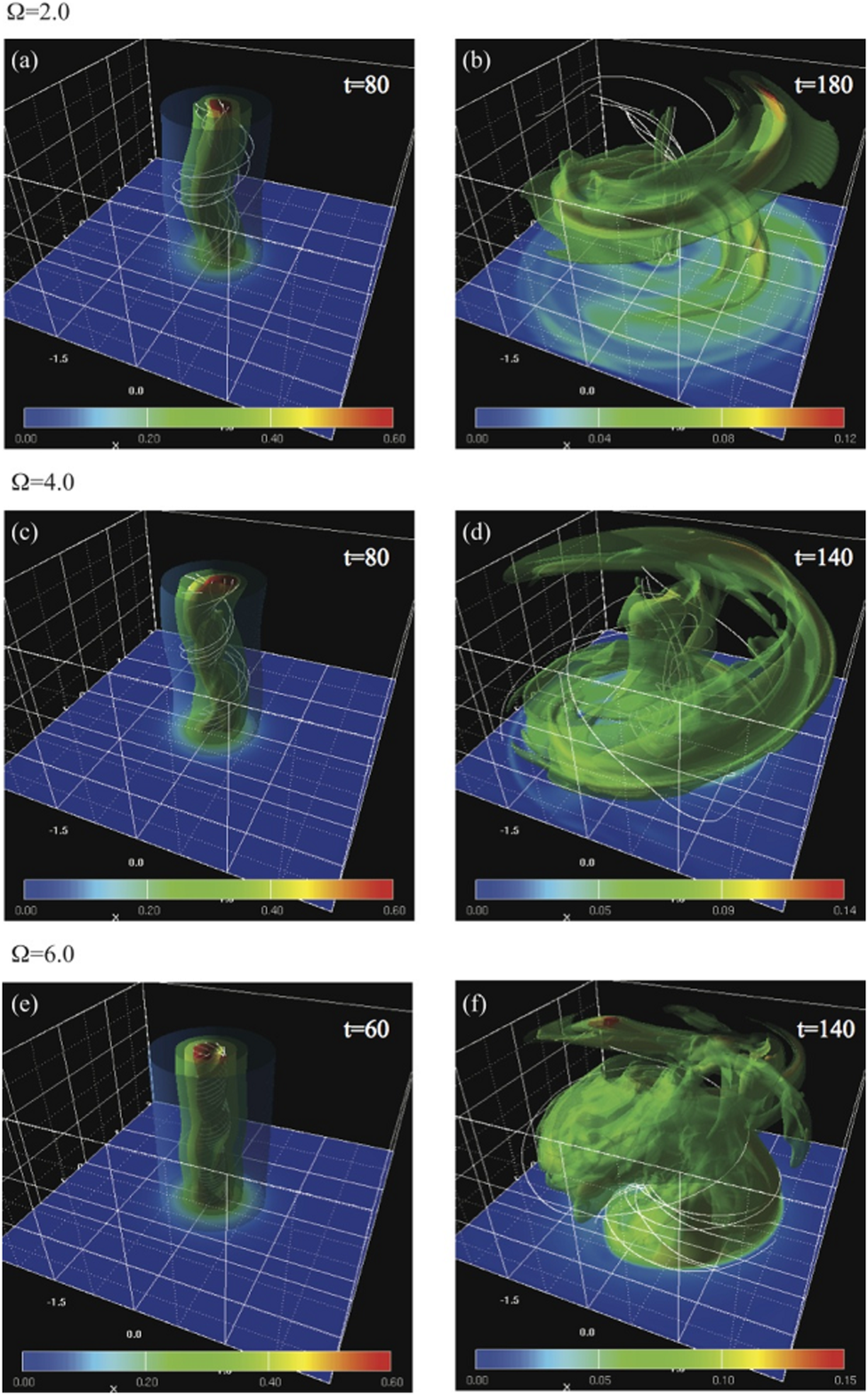}
\caption{Three-dimensional density isosurface with a transverse slice at $z=0$ with constant helical pitch for $\Omega_{0}=2$ (a,b), $4$ (c,d), and $6$ (e,f).  Color shows the logarithm of the density with solid magnetic field lines. In this figure, the dependence of the global structure of instability on the angular rotation is shown.
\label{3D_CPO2-6L}}
\end{figure}
\begin{figure}[h!]
\epsscale{0.7}
\plotone{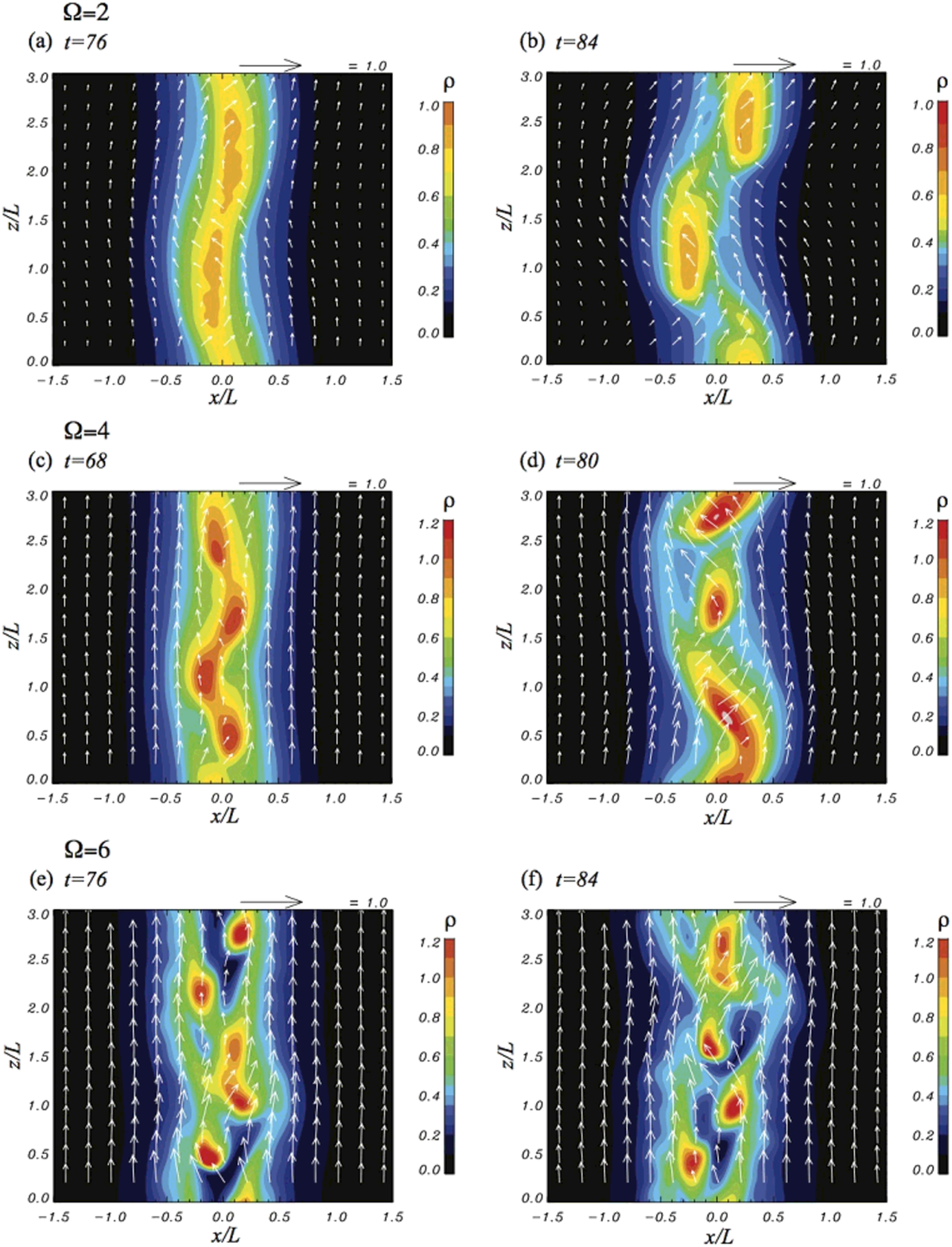}
\caption{Time evolution of two-dimensional images of density in the xz-plane at $y=0$ for $\Omega_{0}=2$ (a,b), $4$ (c,d), and $6$ (e,f) with $\alpha=1$.  Arrows indicate the poloidal velocity. \label{2Dxz_ro_alp1O2-6}}
\end{figure}
In the $\Omega_{0}=2$ case, the behavior of the growing kink in the linear and nonlinear phase is very similar to that in the $\Omega_{0}=1$ case. In both cases, only the $n=1$ kink mode wavelength grows. The propagation speed of the kink is $\sim 0.15c$ and slightly slower than the maximum axial drift speed of $\lesssim 0.25c$.
In the $\Omega_0=4$ case, both $n=1$ and $n=2$ kink mode wavelengths grow (see Fig. 7c and 8c). This is because the pitch decreases with the increasing $\Omega$ and the shorter $n=2$ wavelength is now unstable. In the nonlinear phase, only the $n=1$ kink mode wavelength is excited far from the axis where the pitch is larger. The propagation speed of the kink is $\sim 0.35c$ and the initial maximum axial drift speed is $\lesssim 0.5c$.
In the case $\Omega_{0}=6$, one can also see $n=1$ and $n=2$ kink mode wavelengths growing. In the nonlinear phase, growth of the CD kink instability produces a complicated radially expanding structure as a result of the coupling of multiple wavelengths, and the cylindrical jet structure is almost disrupted. This indicates that taking into account the coupling of multiple unstable wavelengths is crucial to determining whether the jet is eventually disrupted. In this case the propagation speed of the kink is $\sim 0.45c$, and the initial maximum axial drift speed is $\sim 0.6c$.

In order to investigate multiple wavelength growth and coupling in moderate angular velocity cases ($\Omega_0=2$ and $4$), we have performed simulations using longer simulation box ($L_{z}=5L$ for $\Omega_0=2$ and $4L$ for $\Omega_0=4$) and the results are shown in Figure 9.
\begin{figure}[h!]
\epsscale{0.75}
\plotone{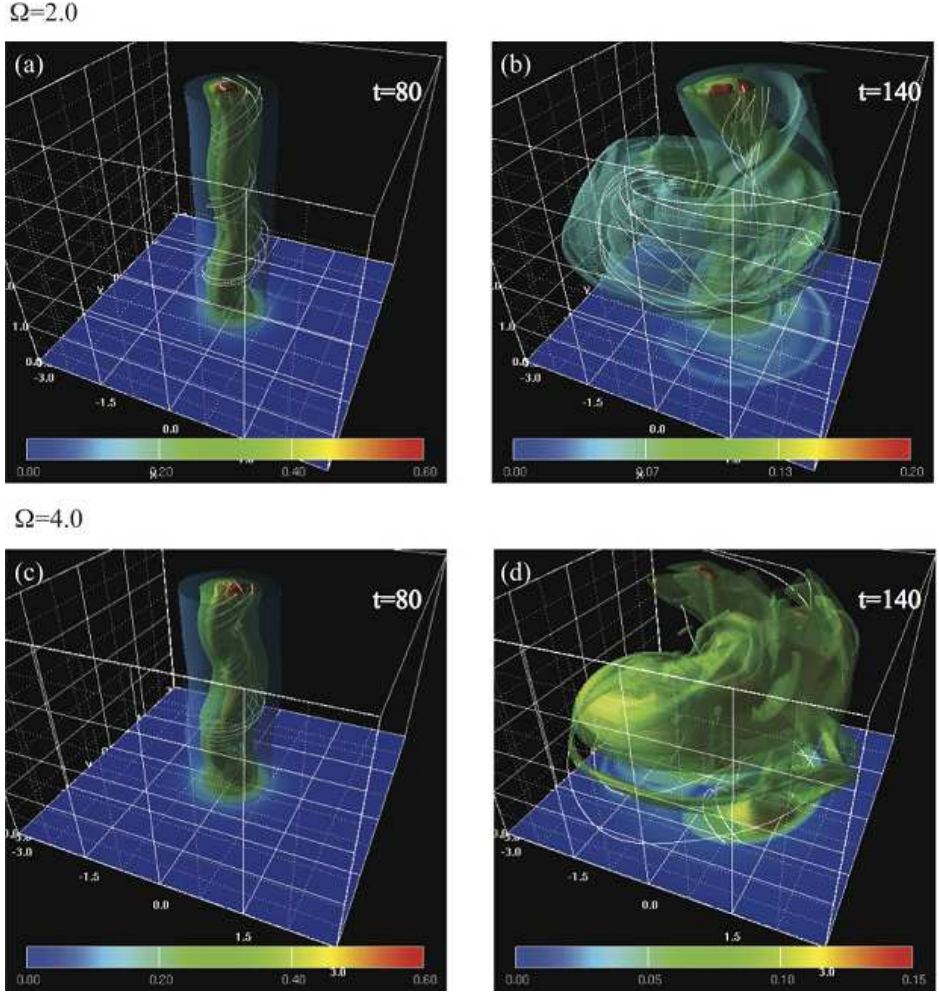}
\caption{Three-dimensional density isosurface with a transverse slice at $z=0$ for $\alpha=1$ with $\Omega_{0}=2$ (a,b), and $4$ (c,d) in a longer simulation box (cases {\it lg}). Color shows the logarithm of the density with solid magnetic field lines.
\label{3D_CPO2-4L1}}
\end{figure}
In this longer simulation box we now see development of both the $n = 1$ and $n=2$ kink mode wavelengths in the $\Omega_0=2$ simulation where in the shorter box only the $n = 1$ kink mode wavelength was observed. Note that the $n=1$ and $n=2$ wavelengths are intrinsically longer in the longer simulation box. Growth of the multiple wavelengths leads to a considerably different structure at longer times as a result of  multiple wavelength interaction. At least initially the result of the $\Omega_0=4$ simulation in the longer simulation box looks similar to that of the $\Omega_0=4$ simulation in the shorter simulation box. However, the structure is again considerably different at longer times even though two wavelengths were excited initially in the shorter box.  Here the difference occurs because the shorter simulation box only allowed the $n=2$ wavelength to grow near to the axis. These differences serve to show how important the coupling of multiple wavelengths can be to the long term development of the instability in the non-linear stage, and we see that
 coupling between multiple wavelengths can lead to disruption of cylindrical jet structure in the nonlinear stage.

\subsection{Dependence on transverse poloidal magnetic field distribution}

According to the linear theory of the kink instability in relativistic force-free jets, the instability growth rate decreases with decreasing outward gradient of the poloidal magnetic field and becomes zero when the poloidal field is homogeneous (Istomin \& Pariev 1996; Lyubarskii 1999).
In order to study the influence of the transverse poloidal magnetic field profile, we have performed simulations with four different  poloidal field profile parameters, $\alpha$. We choose $\alpha \le 1$ which leads to increase in the pitch, P, because it is expected that in jets, the magnetic field lines are less tightly wrapped with radius.

A density isosurface for  $\alpha=0.75$, $0.5$, and $0.35$ with $\Omega_0=4$  is shown in Figure 10.
\begin{figure}[hp]
\epsscale{0.75}
\plotone{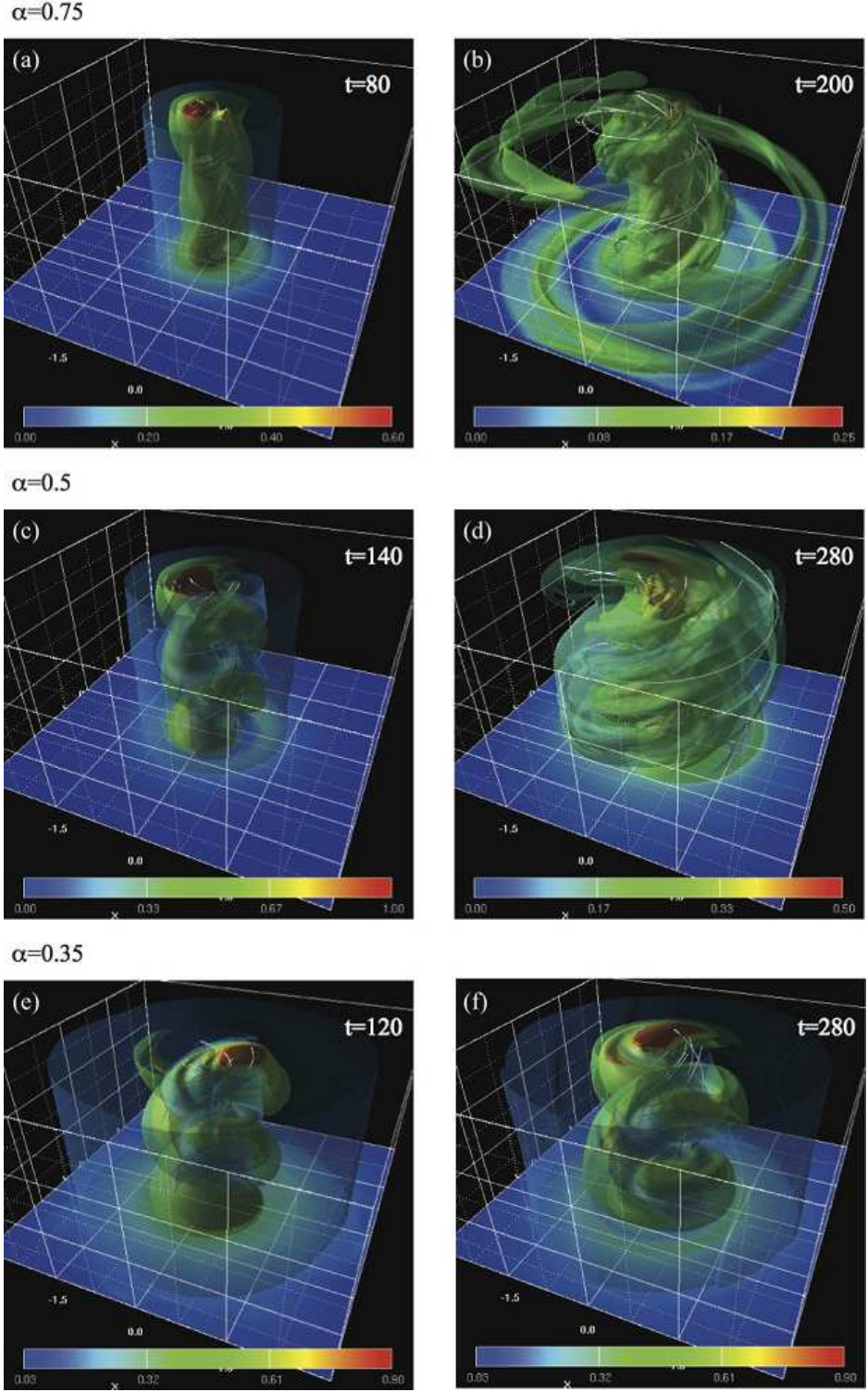}
\caption{Three-dimensional density isosurface with a transverse slice at $z=0$ for $\alpha=0.75$ (a,b), $0.5$ (c,d), and $0.35$ (e,f) with $\Omega_0=4$ and $L_z=3 L$. Color shows the logarithm of the density with solid magnetic field lines. In this figure, the dependence of the structure developed by the instability on the transverse poloidal magnetic field distribution (determined by the parameter $\alpha$ from eq. 2) is shown.
\label{3D_IPomg4S}}
\end{figure}
For $\alpha=0.75$, a helical density structure with the $n=2$ kink mode wavelength develops near the axis by $t_{c}=60$.  In the nonlinear phase, some expanding helical density structure associated with the $n=1$ wavelength is seen at large radius but near the axis, cylindrical density structure remains. The evolution is similar to the $\alpha = 1$ case but the evolution time is longer.  For $\alpha=0.5$, helically twisted density structure with the $n=2$ wavelength is clearly seen around the transition from the linear to the nonlinear phase. In the nonlinear phase, the helical density structure slowly expands radially. For $\alpha=0.35$, helically twisted density structure does not expand significantly in the nonlinear phase. Thus our simulations indicate that for smaller $\alpha$, the CD kink instability can grow only near axis and non-linear growth is suppressed. We note that smaller $\alpha$ cases have stronger magnetic fields and faster initial rotation.  The faster rotation near the axis creates some numerical noise near the axis at early simulation times. However, this numerical noise does not affect our simulation results.

Figure 11 shows the evolution of the volume-averaged relativistic energies determined within a cylinder of radius $R/L \le 2.0 ~(R \le 8R_{0})$ for $\alpha=0.75$, $0.5$, and $0.35$ with $\Omega_0=4$.
\begin{figure}[h!]
\epsscale{0.85}
\plotone{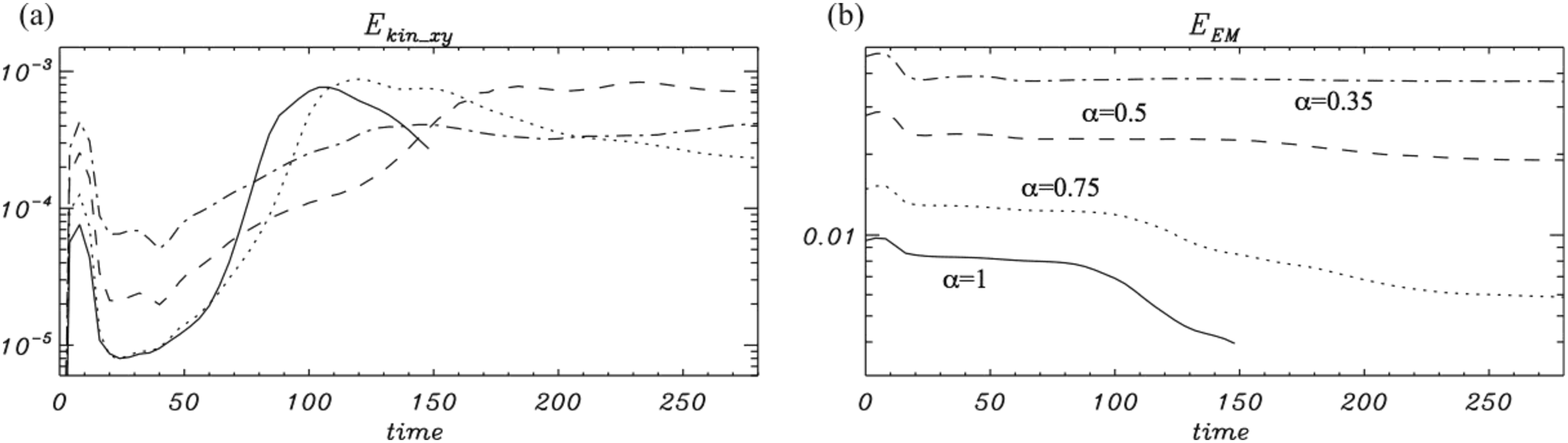}
\caption{Time evolution of (a) $E_{kin,xy}$ and (b) $E_{EM}$ for $\alpha=1.0$ (solid line), $0.75$ (dotted line), $0.5$ (dashed line), and $0.35$ (dash-dotted line) with $\Omega_0=4$. In this figure, the dependence of the instability on the transverse poloidal magnetic field distribution is shown. \label{tev_dp1alp1a}}
\end{figure}
For $\alpha=0.75$ (dotted lines), the evolution profile is similar to that of the $\alpha=1$ case although the scaling is different. Compared to the $\alpha=1$ case, the transition from the linear to the nonlinear phase occurs slightly later ($ t_{c} \simeq 105$). For $\alpha=0.5$ (dashed lines), the growth rate is much lower than for the $\alpha=1$ case. The linear growth phase lasts until $t_{c} \simeq 160$. In the nonlinear phase, kinetic and magnetic energies do not decrease. This indicates non-linear stabilization and structure developed initially is maintained in the nonlinear stage. For $\alpha=0.35$ (dash-dotted lines), transition to the nonlinear phase happens at a lower maximum value of $E_{kin, xy}$ than for the $\alpha=0.5$ case. The growth rate in the $\alpha=0.35$ case is lower than that in the $\alpha=0.5$ case.
These simulations with different transverse poloidal magnetic field distribution, show that the transverse poloidal magnetic field distribution strongly affects growth of the CD kink instability. In accordance with the linear theory, the instability grows slower with decreasing outward gradient of the poloidal field and the simulations show that the instability can saturate in this case.

\section{Summary and Discussion}

We have investigated the influence of rotation and differential motion on the linear and nonlinear development of the CD kink instability of Poynting dominated jets in 3D RMHD simulations. This work provides an extension of our previous study using a static column (Mizuno et al. 2009), and both studies can be considered to represent CD kink instability development in a comoving frame. Our combined studies show that results are strongly affected by the choice of the transverse gas density and pressure profile where the constant gas density and pressure in the previous static column simulations led to matter domination in the periphery where the magnetic field is smaller than in the core. For this reason, the growth of the CD kink instability was suppressed in the nonlinear phase in the earlier static simulations. In the present paper, we have taken the gas density and the pressure to decrease outwards together with the magnetic field so that the system is magnetically dominated everywhere.

According to a linear analysis of the stability of relativistic force-free jets (Istomin \& Pariev 1996; Lyubarskii 1999), the growth rate of the kink instability decreases with the decreasing gradient of the poloidal magnetic field and becomes zero if the poloidal field is homogeneous. This is partially a relativistic effect because in non-relativistic force-free flows, for the hoop stress to be balanced by the the poloidal field requires a gradient in the poloidal field and the poloidal field cannot be constant (with the exception of singular cases like $B_{\phi}\propto 1/r$).  In the relativistic case, the hoop stress can also be partially balanced by the electric force (LHS of eq. (1)) and thus the poloidal field gradient can be smaller than in the non-relativistic case or even vanish.

Our simulations confirm the strong dependence of the instability on the lateral distribution of the poloidal magnetic field; the last is controlled, in our setup, by the parameter $\alpha$ (see eq. 2). For $\alpha\sim 1$, which corresponds to a significant decrease of the poloidal field in the jet core, unstable perturbations grow until the initial cylindrical structure is disrupted, with the disruption caused by the non-linear interaction of multiple growing wavelengths. Note that the unstable perturbations match the helical structure of the magnetic field lines so the wavelength of unstable perturbations is determined by the magnetic pitch. In relativistic jet flows, the toroidal field can significantly exceed the poloidal one (because of the presence of the electric force), the pitch can be much less than the jet radius, and multiple wavelengths are inevitably excited. However, we find that a decrease in $\alpha$ and accompanying more homogeneous poloidal field leads to slower perturbation growth in accord with linear stability analysis, non-linear saturation of the instability, and the initial cylindrical structure is not disrupted.

Our simulations were performed in the frame comoving with the jet. In a differentially moving flow,  any frame in which the velocities are mildly relativistic may be considered as a comoving frame. The instability develops for a time $\tau\sim 100 R_0/c$ in the comoving frame therefore in the lab frame, the instability develops at a scale $z\sim 100\gamma R_0$, where $\gamma$ is the jet Lorentz factor. The instability could develop only if the Alfv\'en crossing time is less than the proper propagation time,
\begin{equation}
R_0 < z/\gamma.
\end{equation}
One can show (Lyubarsky 2009) that such jets should be narrow enough, $R_0 < \sqrt{cz/\Omega}$, and
at any distance from the source, the structure of the flow in these jets is the same as the structure of an appropriate cylindrical equilibrium configuration. This justifies our using equilibrium cylindrical configurations as an initial setup for the stability analysis. When propagating through a medium with decreasing pressure jets expand. Condition (12) is satisfied if the external pressure decreases with the distance not faster than $1/z^2$. In this case the flow is accelerated such that  $\gamma\sim \Omega R_0/c$ until the kinetic energy of the flow becomes comparable with the magnetic energy (i.e., until $\sigma$, the ratio of the Poynting to the plasma energy fluxes, drops to about unity). After equipartition is reached, the flow is accelerated extremely slowly so that the true matter dominated stage (when $\sigma$ drops to, say, 0.1) is reached only at unreasonably large distances (Lyubarsky 2010a).

Simulations of jet launching by a spinning accreting black hole reveal that in real Poynting dominated jets, the poloidal field is very close to uniform (Tchekhovskoy et al. 2008). According to our results, the CD kink instability only weakly develops in this case. This explains why McKinney \& Blandford (2009) did not observe the kink instability in their simulations. However, when the jet is accelerated up to $\sigma\sim 1$, the poloidal flux is concentrated towards the axis (Beskin \& Nokhrina 2009; Lyubarsky 2009). We have shown that such a configuration is subject to disruptive kink instability and therefore our study suggests that Poynting dominated jets are accelerated up to approximate equipartition between the kinetic and magnetic energies and then the regular structure is disrupted by the kink instability. Since the jet is already highly relativistic at this stage, the flow remains collimated but the magnetic energy is efficiently dissipated.

Recall that such a scenario is valid only for jets satisfying $\gamma\theta<1$, where $\theta$ is the opening angle of the jet. This condition is rather restrictive; for example it is violated in gamma-ray bursts (e.g. Tchekhovskoy et al. 2010). In AGNs, the observations show a large scatter in $\gamma\theta$, the median value being $\langle\gamma\theta\rangle=0.26$ (Pushkarev et al. 2009).
One can speculate that the violent magnetic dissipation triggered by the CD kink instability gives rise to the flaring activity of blazars. The observations suggest that the flares can occur relatively far from the central engine, at the distance of hundreds and thousands gravitational radii (Tavecchio et al. 2011). This agrees with our conjecture that the jet is first accelerated till $\sigma\sim 1$ and only then does the kink instability come into play.

The effect of jet rotation considered in the present instability study provides a more realistic setup than that provided by a static plasma column (rigidly moving flow). However, we still have not included the lateral spatial expansion expected for relativistic jets. Jet expansion has strong consequences for the behavior of instabilities on jets modeled as cylinders of constant radius. Jet expansion has  a stabilizing effect, since magnetic instabilities become ineffective when the Alfv\'{e}n speed drops below the lateral expansion speed of the jet. Currently only a few non-relativistic MHD simulations have explored the  effect of jet expansion on the development of CD kink instability (Moll et al. 2008; Moll 2009). In future work, we plan to investigate the effect of jet expansion on the development of CD kink instability in the relativistic MHD regime.

In our current and past simulation work, we assume the ideal MHD condition. Thus, relativistic magnetic reconnection (e.g., Blackman \& Field 1994; Lyubarsky 2005; Zenitani et al. 2010b; Takahashi et al. 2011; Zanotti \& Dumbser 2011) cannot be treated. Development of resistive relativistic MHD (RRMHD) codes are now one of the frontiers of relativistic MHD (e.g., Watanabe \& Yokoyama 2006; Komissarov 2007; Palenzuela et al. 2009; Dumbser \& Zanotti 2009; Zenitani et al. 2010b; Takahashi et al. 2011; Takamoto \& Inoue 2011). In future work, we plan to investigate relativistic magnetic reconnection triggered by the CD kink instability.

\acknowledgments

This work has been supported by NSF awards AST-0506719, AST-0506666, AST-0908010, and AST-0908040, NASA awards NNG05GK73G, NNX07AJ88G, and NNX08AG83G, and US-Israeli BSF award 2006170. Y.M. also acknowledges partially support from Taiwan National Science Council under the grant NSC 100-2112-M-007-022-MY3 and from NAOJ Visiting Scholar Program (Short-term). Y.L. also acknowledges support from Israeli Science Foundation under the grant 737/07. The simulations were performed on the Columbia Supercomputer at the NAS Division of the NASA Ames Research Center, the SR16000 at YITP in Kyoto University, the SGI Altix (ember) at the National Center for Supercomputing Applications in the TeraGrid project and the Nautilus at the National Institute for Computational Sciences in the XSEDE project supported by National Science Foundation.

\end{document}